\documentclass[conference]{IEEEtran}
\IEEEoverridecommandlockouts
\usepackage{amsmath,amsfonts}
\usepackage{algorithm}
\usepackage{algpseudocode}
\usepackage{array}
\usepackage{tikz}
\usetikzlibrary{positioning}
\usepackage[caption=false,font=footnotesize,labelfont=sf,textfont=sf]{subfig}
\usepackage[colorlinks=true, linkcolor=red, citecolor=red, urlcolor=blue]{hyperref}
\usepackage{textcomp}
\usepackage{stfloats}
\usepackage{diagbox}
\usepackage{url}
\usepackage{verbatim}
\usepackage{graphicx}
\usepackage{cite}
\usepackage{bm}
\usepackage{booktabs}
\usepackage{tabularx}
\usepackage{tikz}
\usetikzlibrary{positioning,arrows.meta,fit,calc,shapes.geometric}
\usepackage{amsthm}
\usepackage{proof}
\usepackage{amssymb}
\usepackage{color}
\usepackage{dsfont} 

\usepackage[most]{tcolorbox}
\definecolor{waraNavy}{HTML}{08236C}
\definecolor{waraBlue}{HTML}{0D5BCB}
\definecolor{waraLane}{HTML}{F2F7FF}
\definecolor{waraSideTop}{HTML}{225BAD}
\definecolor{waraSideBottom}{HTML}{153F86}
\definecolor{waraGreen}{HTML}{EAF7EF}
\definecolor{waraGreenLine}{HTML}{16803A}
\definecolor{waraTag}{HTML}{EEF5FF}
\definecolor{waraTagLine}{HTML}{7DB1FF}
\definecolor{waraOrange}{HTML}{9B4D00}
\definecolor{waraRepair}{HTML}{C0392B}

\usepackage{etoolbox}
\makeatletter
\patchcmd{\@makecaption}
{{\normalfont\footnotesize\scshape #2}}
{{\normalfont\footnotesize #2}}
{}{}
\makeatother
\newtcolorbox{warabox}[1]{
	breakable,
	colback=gray!3,
	colframe=black!55,
	boxrule=0.45pt,
	arc=1pt,
	left=5pt,
	right=5pt,
	top=5pt,
	bottom=5pt,
	fonttitle=\bfseries\footnotesize,
	fontupper=\scriptsize,
	coltitle=black,
	colbacktitle=gray!18,
	title={#1},
	before skip=6pt,
	after skip=6pt
}

\definecolor{litPurpleLine}{HTML}{7B5DD6}
\definecolor{litRedLine}{HTML}{C0392B}

\allowdisplaybreaks
\def\BibTeX{{\rm B\kern-.05em{\sc i\kern-.025em b}\kern-.08em
		T\kern-.1667em\lower.7ex\hbox{E}\kern-.125emX}}

\begin{document}
	
	\title{WARA: A Closed-Loop Multi-Agent Framework for Wireless Optimization Autoresearch
		\thanks{The work was supported in part by the Guangdong Major Project of Basic and Applied Basic Research (No. 2023B0303000001), the Basic Research Project No. HZQB-KCZYZ-2021067 of Hetao Shenzhen-HK S\&T Cooperation Zone, the National Natural Science Foundation of China under grants Nos. U25A20390 and 62471424, the Guangdong Provincial Key Laboratory of Future Networks of Intelligence under grant No. 2022B1212010001, and the Shenzhen Fundamental Research Program under grant No. JCYJ20250604141209012.}
	}
	
	\author{
		\IEEEauthorblockN{
			Yuan Guo$^{\dagger \S}$,
			Yilong Chen$^{\dagger \S}$,
			Chao Hu$^{\dagger \S}$,
			Xianghao Yu$^{\ddagger}$,
			Liang Hong$^{\P}$,
			and Jie Xu$^{\dagger \S}$
		}
		\IEEEauthorblockA{
			$^{\dagger}$School of Science and Engineering (SSE), The Chinese University of Hong Kong (Shenzhen), Shenzhen, China\\
			$^{\S}$Shenzhen Future Network of Intelligence Institute (FNii-Shenzhen), Shenzhen, China\\
			$^{\ddagger}$Department of Electrical Engineering, City University of Hong Kong, Hong Kong, China\\
			$^{\P}$Sun Yat-Sen University, Guangzhou, China\\
			Emails: guoyuan@cuhk.edu.cn, chenyilong@cuhk.edu.cn, chaohu@link.cuhk.edu.cn,\\
			alex.yu@cityu.edu.hk, hongliang@sysu.edu.cn, xujie@cuhk.edu.cn
		}
	}
	\maketitle

	\begin{abstract}
		Large language model (LLM) agents have shown growing capabilities in tool use, code execution, artifact inspection, and iterative revision, creating new opportunities for automating scientific research. To the best of our knowledge, this paper presents the first end-to-end autoresearch framework for the wireless domain, with a particular focus on wireless resource allocation optimization, an essential area for characterizing the fundamental performance limits of wireless systems and enhancing their practical performance under dynamic channel and network conditions. Specifically, we propose the Wireless AutoResearch Agent (WARA), a closed-loop multi-agent system that transforms an initial research topic into a complete research package. WARA organizes the research workflow into three phases: 1) research gap identification and problem proposal, 2) optimization modeling, algorithm design, and experimentation, and 3) research deliverable construction. Each phase follows an artifact-mediated process, in which structured upstream artifacts are consumed to generate downstream outputs. Controller-managed gates validate these artifacts and maintain consistency among problem formulations, algorithms, experiments, and research claims. When validation fails, WARA repairs only the affected artifact instead of restarting the entire workflow. We further design an LLM-based ScoringAgent to evaluate manuscript-level research validity. Comparative results show that WARA substantially outperforms one-shot LLM generation and approaches the quality profile of recently accepted peer-reviewed papers. These results demonstrate the potential of closed-loop artifact control for end-to-end LLM-assisted wireless optimization research. The source code is available at \url{https://github.com/guoyuan-dotcom/WARA_CUHKSZ}.
	\end{abstract}


	\section{Introduction}
	
	The remarkable success of large artificial intelligence models, particularly large language models (LLMs), has reshaped expectations regarding the role of artificial intelligence (AI) in scientific and engineering research. Beyond their established use in text and code generation, LLMs are increasingly being embedded into agentic systems that integrate language-based reasoning with external tools, executable environments, reflection mechanisms, and multi-agent coordination ~\cite{yao2023react,schick2023toolformer,shinn2023reflexion,wu2023autogen}. In this context, LLM-based agentic systems offer a promising path toward autonomous research exploration, enabling the generation of innovative research ideas and their progressive refinement into validated research outcomes through iterative reasoning, execution, and verification.
	
	A typical research workflow in science and engineering can generally be divided into three phases: 1) research gap and problem identification; 2) problem solving and experimentation; and 3) construction of deliverables. Prior work has shown that LLMs can assist these phases separately, including research-idea generation and problem proposal~\cite{si2024ideas,baek2024researchagent,gottweis2025aicoscientist}, design--execution loops for coding, experimentation, and optimization modeling~\cite{huang2025orlm}, and paper writing and review through iterative research--review cycles~\cite{ifargan2024datatopaper,weng2024cycleresearcher,zhu2025deepreview}. Different from these phase-specific efforts, another line of work connects multiple stages into closed-loop autoresearch workflows. For example, AI Scientist~\cite{lu2024aiscientist} and its successor AI Scientist-v2~\cite{yamada2025aiscientistv2} generate complete papers from research ideas, AI-Researcher~\cite{tang2025airesearcher} builds a multi-agent system for autonomous AI research, and AutoResearchClaw~\cite{autoresearchclaw2026} incorporates failure recovery and result verification, while Robin~\cite{ghareeb2026multi} extends the paradigm to biology with human experimentation in the loop.
	
	Motivated by these advances, this paper investigates closed-loop research automation for wireless communications, with a particular focus on wireless optimization. This domain provides a natural starting point because many fundamental wireless problems, including those arising in multiple-input multiple-output (MIMO) communications~\cite{1266912PAULRAJ}, reconfigurable intelligent surface (RIS)-assisted networks~\cite{wu2019intelligent}, and integrated sensing and communications (ISAC) systems~\cite{Liu9737357}, can be expressed as mathematical programs involving signal design, radio-resource allocation, system constraints, and performance objectives. More importantly, wireless optimization research follows a structured and executable workflow: researchers formulate mathematical problems from channel models, performance requirements, and resource constraints, develop analytical or numerical solution methods, and evaluate the resulting designs through simulation~\cite{tse2005fundamentals,goldsmith2005wireless}. This workflow provides natural verification signals, as optimization solvers and simulators can be repeatedly invoked to assess the feasibility, correctness, and consistency of the generated models, algorithms, experimental results, and technical claims.

	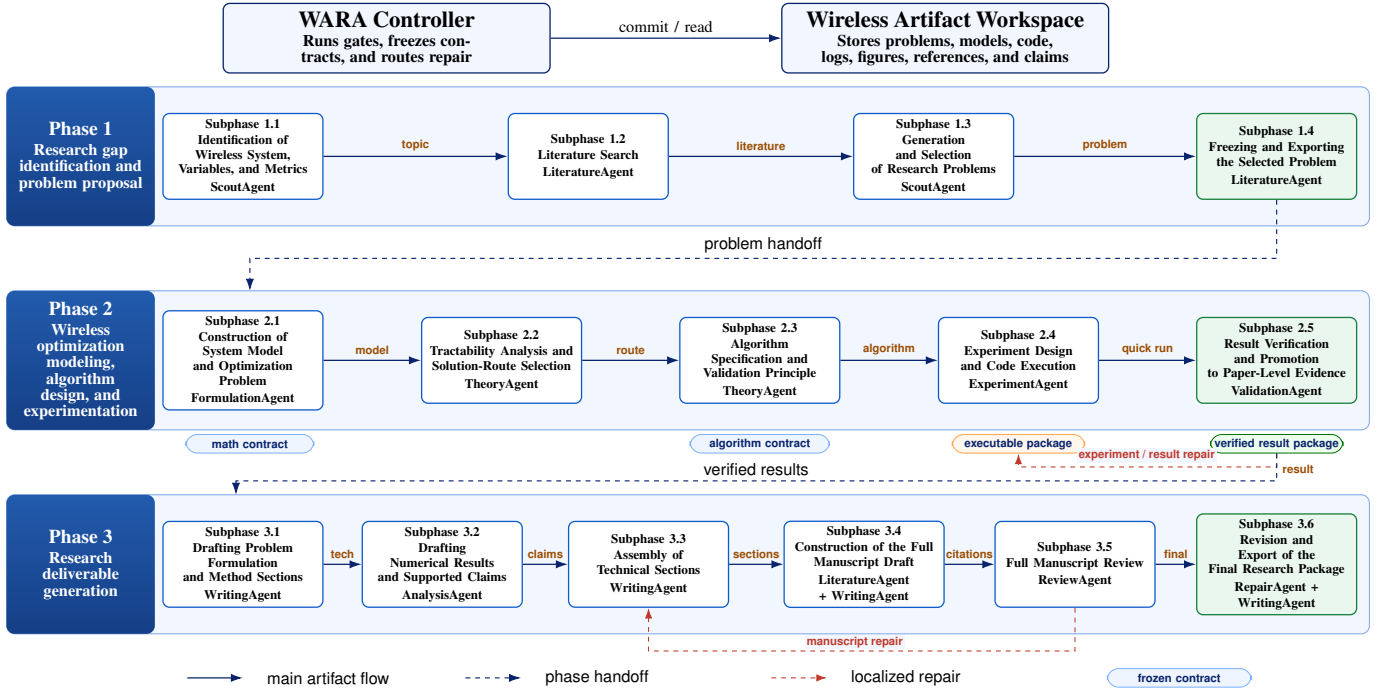
\begin{figure*}[!t]
		\centering
		\resizebox{\textwidth}{!}{%
			\begin{tikzpicture}[
				x=1cm,
				y=1cm,
				font=\small,
				flow/.style={-{Latex[length=2.6mm,width=1.8mm]}, draw=waraNavy, line width=0.78pt},
				handoff/.style={-{Latex[length=2.6mm,width=1.8mm]}, draw=waraNavy, dashed, dash pattern=on 3pt off 3pt, line width=0.70pt},
				repair/.style={-{Latex[length=2.3mm,width=1.6mm]}, draw=waraRepair, dashed, dash pattern=on 3pt off 3pt, line width=0.65pt},
				lane/.style={draw=waraBlue!55, rounded corners=5pt, line width=0.56pt, fill=waraLane},
				phasebox/.style={draw=waraBlue, rounded corners=4pt, line width=0.80pt, fill=white, align=center, text width=3.10cm, minimum width=3.45cm, minimum height=1.85cm, inner sep=3pt},
				frozenstep/.style={draw=waraGreenLine, rounded corners=4pt, line width=0.80pt, fill=waraGreen, align=center, text width=3.10cm, minimum width=3.45cm, minimum height=1.85cm, inner sep=3pt},
				topbox/.style={draw=waraNavy, rounded corners=4pt, line width=0.80pt, fill=waraLane, align=center, text width=6.80cm, minimum width=7.05cm, minimum height=1.5cm, inner sep=2pt},
				tag/.style={draw=waraTagLine, rounded corners=7pt, line width=0.50pt, fill=waraTag, align=center, minimum height=0.40cm, inner sep=1.5pt},
				taggreen/.style={draw=green!45!black, rounded corners=7pt, line width=0.50pt, fill=green!7, align=center, minimum height=0.40cm, inner sep=1.5pt},
				tagorange/.style={draw=orange!70, rounded corners=7pt, line width=0.50pt, fill=orange!7, align=center, minimum height=0.40cm, inner sep=1.5pt},
				side/.style={rounded corners=5pt, align=center, text=white},
				gatelabel/.style={font=\sffamily\bfseries\scriptsize, text=waraOrange},
				repairlabel/.style={font=\sffamily\bfseries\scriptsize, text=waraRepair, fill=waraLane, inner sep=1pt},
				tagtext/.style={font=\sffamily\bfseries\scriptsize, text=waraNavy}
				]
				
				\node[topbox] (controller) at (8.70,14)
				{\Large\bfseries WARA Controller\\[0.5mm]
					\normalsize Runs gates, freezes contracts, and routes repair};
				
				\node[topbox] (workspace) at (20.75,14)
				{\Large\bfseries Wireless Artifact Workspace\\[0.5mm]
					\normalsize Stores problems, models, code, logs, figures, references, and claims};
				
				\draw[flow] (controller.east) -- node[above, font=\normalsize] {commit / read} (workspace.west);
				
				\draw[lane] (0.50,9.95) rectangle (29.9,12.95);
				\draw[lane] (0.50,5.55) rectangle (29.9,8.55);
				\draw[lane] (0.50,1.15) rectangle (29.9,4.15);
				
				\shade[rounded corners=5pt, top color=waraSideTop, bottom color=waraSideBottom]
				(0.50,9.95) rectangle (3.70,12.95);
				\shade[rounded corners=5pt, top color=waraSideTop, bottom color=waraSideBottom]
				(0.50,5.55) rectangle (3.70,8.55);
				\shade[rounded corners=5pt, top color=waraSideTop, bottom color=waraSideBottom]
				(0.50,1.15) rectangle (3.70,4.15);
				
				\node[side, text width=3.05cm] at (2.10,11.45)
				{\large\bfseries Phase 1\\[0.8mm]\normalsize\bfseries Research gap identification and problem proposal};
				
				\node[side, text width=3.05cm] at (2.10,7.05)
				{\large\bfseries Phase 2\\[0.8mm]\normalsize\bfseries Wireless optimization modeling, algorithm design, and experimentation};
				
				\node[side, text width=3.05cm] at (2.10,2.65)
				{\large\bfseries Phase 3\\[0.8mm]\normalsize\bfseries Research deliverable generation};
				
				\node[phasebox] (p11) at (5.6,11.45)
				{\footnotesize\bfseries Subphase 1.1\\[0.3mm]
					Identification of Wireless System,\\Variables, and Metrics\\[0.3mm]
					{\footnotesize ScoutAgent}};
				
				\node[phasebox] (p12) at (13.05,11.45)
				{\footnotesize\bfseries Subphase 1.2\\[0.3mm]
					Literature Search\\[0.3mm]
					{\footnotesize LiteratureAgent}};
				
				\node[phasebox] (p13) at (20.50,11.45)
				{\footnotesize\bfseries Subphase 1.3\\[0.3mm]
					Generation and Selection\\of Research Problems\\[0.3mm]
					{\footnotesize ScoutAgent}};
				
				\node[frozenstep] (p14) at (27.9,11.45)
				{\footnotesize\bfseries Subphase 1.4\\[0.3mm]
					Freezing and Exporting\\the Selected Problem\\[0.3mm]
					{\footnotesize LiteratureAgent}};
				
				\draw[flow] (p11.east) -- node[above, gatelabel] {topic} (p12.west);
				\draw[flow] (p12.east) -- node[above, gatelabel] {literature} (p13.west);
				\draw[flow] (p13.east) -- node[above, gatelabel] {problem} (p14.west);
				
				\draw[handoff] (p14.south) -- (27.9,9.25)
				-- node[above, font=\sffamily\normalsize] {problem handoff}
				(5.75,9.25) -- (5.75,8.55);
				
				\node[phasebox] (p21) at (5.6,7.05)
				{\footnotesize\bfseries Subphase 2.1\\[0.3mm]
					\footnotesize Construction of System Model\\and Optimization Problem\\[0.3mm]
					{\footnotesize FormulationAgent}};
				
				\node[phasebox] (p22) at (11.18,7.05)
				{\footnotesize\bfseries Subphase 2.2\\[0.3mm]
					\footnotesize Tractability Analysis and\\Solution-Route Selection\\[0.3mm]
					{\footnotesize TheoryAgent}};
				
				\node[phasebox] (p23) at (16.75,7.05)
				{\footnotesize\bfseries Subphase 2.3\\[0.3mm]
					\footnotesize Algorithm Specification and\\Validation Principle\\[0.3mm]
					{\footnotesize TheoryAgent}};
				
				\node[phasebox] (p24) at (22.33,7.05)
				{\footnotesize\bfseries Subphase 2.4\\[0.3mm]
					\footnotesize Experiment Design\\and Code Execution\\[0.3mm]
					{\footnotesize ExperimentAgent}};
				
				\node[frozenstep] (p25) at (27.9,7.05)
				{\footnotesize\bfseries Subphase 2.5\\[0.3mm]
					\footnotesize Result Verification and Promotion\\to Paper-Level Evidence\\[0.3mm]
					{\footnotesize ValidationAgent}};
				
				\draw[flow] (p21.east) -- node[above, gatelabel] {model} (p22.west);
				\draw[flow] (p22.east) -- node[above, gatelabel] {route} (p23.west);
				\draw[flow] (p23.east) -- node[above, gatelabel] {algorithm} (p24.west);
				\draw[flow] (p24.east) -- node[above, gatelabel] {quick run} (p25.west);
				
				\node[tag, tagtext, text width=2.65cm] at (5.75,5.25) {math contract};
				\node[tag, tagtext, text width=2.90cm] at (16.75,5.25) {algorithm contract};
				\node[tagorange, tagtext, text width=2.75cm] (xpack) at (22.33,5.25) {executable package};
				\node[taggreen, tagtext, text width=2.75cm] (evtag) at (27.9,5.25) {verified result package};
				
				\draw[repair] (evtag.south) -- (27.9,4.75) -- (22.33,4.75) -- (xpack.south);
				\node[repairlabel] at (25.10,5.00) {experiment / result repair};
				
				\draw[handoff] (evtag.south) -- node[right, yshift=-1pt, gatelabel] {result}
				(27.9,4.45) -- node[above, font=\sffamily\normalsize] {verified results}
				(5.45,4.45) -- (5.45,4.15);
				
				\node[phasebox] (p31) at (5.6,2.65)
				{\footnotesize\bfseries Subphase 3.1\\[0.3mm]
					\footnotesize Drafting Problem Formulation\\and Method Sections\\[0.3mm]
					{\footnotesize WritingAgent}};
				
				\node[phasebox] (p32) at (9.90,2.65)
				{\footnotesize\bfseries Subphase 3.2\\[0.3mm]
					\footnotesize Drafting Numerical Results\\and Supported Claims\\[0.3mm]
					{\footnotesize AnalysisAgent}};
				
				\node[phasebox] (p33) at (14.35,2.65)
				{\footnotesize\bfseries Subphase 3.3\\[0.3mm]
					\footnotesize Assembly of\\Technical Sections\\[0.3mm]
					{\footnotesize WritingAgent}};
				
				\node[phasebox] (p34) at (19,2.65)
				{\footnotesize\bfseries Subphase 3.4\\[0.3mm]
					\footnotesize Construction of the Full\\Manuscript Draft\\[0.3mm]
					{\footnotesize LiteratureAgent + WritingAgent}};
				
				\node[phasebox] (p35) at (23.55,2.65)
				{\footnotesize\bfseries Subphase 3.5\\[0.3mm]
					\footnotesize Full Manuscript Review\\[0.3mm]
					{\footnotesize ReviewAgent}};
				
				\node[frozenstep] (p36) at (27.9,2.65)
				{\footnotesize\bfseries Subphase 3.6\\[0.3mm]
					\footnotesize Revision and Export of the\\Final Research Package\\[0.3mm]
					{\footnotesize RepairAgent + WritingAgent}};
				
				\draw[flow] (p31.east) -- node[above, gatelabel] {tech} (p32.west);
				\draw[flow] (p32.east) -- node[above, gatelabel] {claims} (p33.west);
				\draw[flow] (p33.east) -- node[above, gatelabel] {sections} (p34.west);
				\draw[flow] (p34.east) -- node[above, gatelabel] {citations} (p35.west);
				\draw[flow] (p35.east) -- node[above, gatelabel] {final} (p36.west);
				
				\draw[repair] (p35.south) -- (23.55,0.78) -- (14.35,0.78) -- (p33.south);
				\node[repairlabel] at (18.80,0.98) {manuscript repair};
				
				\draw[flow] (4.4,0.20) -- +(1.2,0);
				\node[anchor=west, font=\sffamily\normalsize] at (6.0,0.20) {main artifact flow};
				
				\draw[handoff] (10.4,0.20) -- +(1.2,0);
				\node[anchor=west, font=\sffamily\normalsize] at (12.0,0.20) {phase handoff};
				
				\draw[repair] (17.0,0.20) -- +(1.2,0);
				\node[anchor=west, font=\sffamily\normalsize] at (18.6,0.20) {localized repair};
				
				\node[tag, tagtext, text width=3.0cm] at (25.8,0.20) {frozen contract};
				
			\end{tikzpicture}
		}
		\caption{WARA workflow for wireless optimization autoresearch. Each subphase is executed by one or more role-specialized agents responsible for specific tasks, as summarized in Table \ref{tab:wara_agents}.}
		\label{fig:wara_pipeline}
	\end{figure*}

	Recent studies have begun exploring the use of LLMs in wireless communications, including LLM-assisted optimization and network control~\cite{lee2026llmresource}, as well as domain-specific agent frameworks and benchmarks such as WirelessAgent~\cite{tong2025wirelessagent}, AgentRAN~\cite{elkael2025agentran}, and ComAgent~\cite{li2026comagent}. While these efforts demonstrate the potential of LLMs for wireless intelligence, they mainly address isolated tasks such as network control, optimization formulation, or benchmark evaluation. However, they do not address the end-to-end automation of wireless optimization research, where problem discovery, mathematical modeling, algorithm design, experimentation, result validation, and manuscript generation must remain aligned within a unified workflow. Such cross-stage alignment is critical, since an error introduced in the problem formulation may propagate into the algorithm, experiments, figures, and final claims if no explicit artifact-level control is imposed.
	
	To bridge this gap, we propose the Wireless AutoResearch Agent (WARA), an end-to-end autoresearch framework for wireless optimization. Rather than generating individual artifacts in isolation, WARA transforms an initial topic into a technically consistent and evidence-supported research package by organizing the process into three phases: 1) research gap identification and problem proposal; 2) wireless optimization modeling, algorithm design, and experimentation; and 3) research deliverable generation. For each phase, we design artifact-mediated control mechanisms in which declared upstream artifacts are consumed to produce structured outputs for downstream use, and a controller-managed gate validates the generated artifacts to ensure consistency among models, algorithms, experiments, and claims. When an artifact fails validation, only that artifact is repaired rather than restarting the entire workflow. We further design a structured LLM-based ScoringAgent to assess manuscript-level research validity. Comparative evaluations show that WARA substantially outperforms one-shot LLM generation while approaching the quality of recently accepted peer-reviewed papers, demonstrating that closed-loop artifact control provides a practical path toward reliable end-to-end LLM-assisted wireless optimization research.

	\begin{table*}[!t]
		\centering
		\caption{Role-Specialized Agents in WARA.}
		\label{tab:wara_agents}
		\setlength{\tabcolsep}{3pt}
		\renewcommand{\arraystretch}{1.18}
		\scriptsize
		\begin{tabular}{p{0.13\linewidth} p{0.30\linewidth} p{0.31\linewidth} p{0.18\linewidth}}
			\hline
			\textbf{Agent} & \textbf{Research role} & \textbf{Input / output artifacts} & \textbf{Execution / verification} \\
			\hline
			
			ScoutAgent 
			& Frames the topic, defines scope, and selects a bounded problem direction. 
			& Input: topic, literature signals. Output: research frame, candidate/selected directions, Phase-1 handoff. 
			& Schema, scope, and feasibility checks. \\
			\hline
			
			LiteratureAgent 
			& Grounds the direction in wireless literature; identifies prior models, baselines, and citations. 
			& Input: research frame, selected direction. Output: evidence pack, reference bank, citation--claim map. 
			& Retrieval; metadata and citation checks. \\
			\hline
			
			FormulationAgent 
			& Converts the direction into a system model and original optimization formulation. 
			& Input: Phase-1 handoff, evidence pack. Output: system model, variables, objective, constraints, math contract. 
			& Symbol audit; equation consistency; \LaTeX{} checks. \\
			\hline
			
			TheoryAgent 
			& Analyzes tractability, selects the reformulation route, and develops the algorithm. 
			& Input: math contract, system model, formulation. Output: tractability analysis, algorithm contract and description. 
			& Contract compliance; convexity audit. \\
			\hline
			
			ExperimentAgent 
			& Builds and executes the experiment package per the frozen contracts. 
			& Input: math/algorithm contracts, experiment blueprint. Output: experiment design, Python/CVXPY code, logs, figures. 
			& Code execution; solver and benchmark checks. \\
			\hline
			
			ValidationAgent 
			& Verifies experiment outputs and promotes valid numerical evidence. 
			& Input: experiment report, figure data, logs. Output: evidence contract, verified figures, benchmark definitions. 
			& Log/figure audit; metric and claim-evidence checks. \\
			\hline
			
			AnalysisAgent 
			& Interprets verified evidence and maps trends to research claims. 
			& Input: evidence contract, verified figures. Output: numerical-results text, claim--evidence map. 
			& Claim--evidence and trend checks. \\
			\hline
			
			WritingAgent 
			& Assembles manuscript text from contracts, evidence, and citations. 
			& Input: contracts, results text, figures, reference bank. Output: technical sections, full-paper preview. 
			& \LaTeX{} compilation; bibliography and notation checks. \\
			\hline
			
			ReviewAgent 
			& Audits the artifact chain and draft for consistency, evidence, and readiness. 
			& Input: full-paper preview, claim map, contracts. Output: review report, critical issues, routing decision. 
			& Cross-artifact, evidence, and citation audit. \\
			\hline
			
			RepairAgent 
			& Applies scoped revisions per gate or review feedback. 
			& Input: target artifact, error report, contract. Output: repaired artifact, repair log, gate rerun result. 
			& Localized repair; regression; gate rerun. \\
			\hline
		\end{tabular}
	\end{table*}

	\section{WARA Framework}
	\label{sec:architecture}
	
	The central idea of WARA is to treat a wireless optimization study as a chain of verifiable research artifacts rather than as a single generated manuscript. Starting from an initial topic, WARA identifies a concrete wireless optimization problem, builds the corresponding model, algorithm, and experiments, and synthesizes the research deliverables from the verified technical results. This artifact-chain view helps maintain alignment among modeling assumptions, optimization objectives, system constraints, solution methods, validation metrics, and technical claims throughout the research process. As illustrated in Fig.~\ref{fig:wara_pipeline}, the workflow is coordinated by a controller that drives a set of role-specialized LLM agents, summarized in Table~\ref{tab:wara_agents}, each responsible for a specific research task such as literature search, formulation, theory analysis, experimentation, writing, or review. Rather than allowing these agents to communicate through free-form text, WARA mediates their interaction through structured artifacts: an artifact denotes any stored research object, such as a formulation, code file, result table, or manuscript section; a contract is an accepted artifact that fixes key technical decisions; and a handoff is the validated package passed between phases.

	The controller governs this artifact chain by storing each generated artifact, applying a validation gate before it is consumed downstream, freezing accepted artifacts as contracts, and routing failed outputs back to the responsible agent for repair. This design makes the research process both consistent and recoverable: because every step is grounded in the upstream contracts it depends on, the model, algorithm, experiments, and claims are kept mutually aligned, and because failures are localized, only the offending artifact is repaired rather than restarting the entire workflow. Following this principle, WARA organizes the workflow into three sequential phases, namely research gap identification and problem proposal (Phase~1), wireless optimization modeling, algorithm design, and experimentation (Phase~2), and research deliverable generation (Phase~3), which are detailed in the following.
	
	\textbf{Phase~1: Research gap identification and problem proposal.} This phase transforms a broad wireless topic into a well-defined optimization problem through four subphases.
	\begin{itemize}
		\item In Subphase~1.1, the initial topic is grounded into a structured research frame. The controller validates the topic against a wireless-domain taxonomy\footnote{The taxonomy covers common wireless tasks, radio technologies, optimization variables, modeling assumptions, solver families, and performance metrics, e.g., ISAC, simultaneous wireless information and power transfer (SWIPT), RIS, beamforming, imperfect channel state information (CSI), successive convex approximation (SCA), semidefinite relaxation (SDR), sum rate, and the Cram\'er--Rao bound (CRB).} to filter out unrelated systems and metrics, after which the ScoutAgent characterizes the wireless scenario, key tradeoffs, optimization variables, candidate objectives, constraints, and evaluation metrics.
		\item In Subphase~1.2, the research frame is grounded in the literature. The LiteratureAgent retrieves prior work from multiple academic sources, including Crossref, Semantic Scholar, OpenAlex, arXiv, and IEEE Xplore, and consolidates the relevant models, baselines, and references into an evidence pack.
		\item In Subphase~1.3, a concrete research problem is selected. The ScoutAgent proposes several candidate optimization problems, from which the controller retains the one that is sufficiently specific, optimization-oriented, technically grounded, and actionable.
		\item In Subphase~1.4, the selected problem is finalized for downstream development. It is assembled with its verified references and modeling instructions into a problem contract, which is frozen as the Phase-1 handoff once it passes the validation gate, and otherwise routed back to the literature-search or problem-selection step for targeted repair.
	\end{itemize}
	
	\textbf{Phase~2: Wireless optimization modeling, algorithm design, and experimentation.} With the Phase-1 handoff frozen, this phase develops the selected problem into a complete technical study through five subphases.
	\begin{itemize}
		\item In Subphase~2.1, the FormulationAgent constructs the wireless system model and optimization problem, specifying the signal and channel models, optimization variables, objective, constraints, and performance metrics. Once the controller verifies that the formulation is complete, physically meaningful, and consistent with the selected problem, it is frozen as the mathematical contract.
		\item In Subphase~2.2, the TheoryAgent analyzes the tractability of the formulated problem by examining convexity, conic representability, and the dominant sources of difficulty, such as nonconvex coupling, discrete variables, or rank constraints, and accordingly selects a solution route, e.g., a convex solver, SDR, SCA, or alternating optimization.
		\item In Subphase~2.3, the TheoryAgent derives an implementable algorithm from the chosen route, specifying the update rules, stopping criteria, and baseline-comparison methodology, which together are frozen as the algorithm contract.
		\item In Subphase~2.4, the ExperimentAgent translates these contracts into an experiment design and the corresponding Python/CVXPY code. The controller then inserts the code into a fixed validation harness\footnote{A validation harness is controller-owned code that standardizes how the generated experiment code is loaded, executed, logged, and checked, providing the execution wrapper and verification layer required for reliable agent execution~\cite{zhong2026aiharness}.} and executes it locally, requiring solver-based methods to invoke the designated solver rather than a proxy implementation.
		\item In Subphase~2.5, the ValidationAgent promotes the execution outputs to paper-level evidence by verifying data provenance, benchmark consistency, metric validity, and figure readiness. A three-level expansion strategy is then applied, progressing from a scout sweep that identifies the intended trend, to a medium-scale sweep that confirms its stability across random seeds, and finally to a paper-scale sweep that produces publication-quality figures, before the verified result package is frozen for Phase~3.
	\end{itemize}
	
	\textbf{Phase~3: Research deliverable generation.} This phase converts the verified technical artifacts into a complete research package through six subphases.
	\begin{itemize}
		\item In Subphase~3.1, the WritingAgent drafts the problem-formulation and method sections from the frozen mathematical and algorithm contracts, ensuring that the notation, equations, and algorithmic procedure remain consistent with the approved contracts.
		\item In Subphase~3.2, the AnalysisAgent drafts the numerical-results section from the verified result package and builds a claim--figure traceability record that links each reported observation to its supporting evidence.
		\item In Subphase~3.3, the WritingAgent assembles the core technical sections, including the abstract, keywords, and conclusion, while preserving consistency with the validated technical content.
		\item In Subphase~3.4, the LiteratureAgent prepares validated BibTeX entries, prioritizing IEEE and peer-reviewed sources, after which the WritingAgent writes the introduction and inserts citations using the verified reference keys to complete the full manuscript draft.
		\item In Subphase~3.5, the ReviewAgent performs a paper-level review of the consistency among equations, algorithms, figures, citations, and technical claims, producing a structured report of blocking, major, and minor issues with corresponding ownership assignments.
		\item In Subphase~3.6, the RepairAgent resolves the flagged issues and the WritingAgent regenerates the manuscript under the guidance of the review report. Once the final quality gate, which validates compilation, citation integrity, figure quality, and claim support, is passed, WARA exports the final research package; otherwise, the issue is routed to local repair or back to the responsible upstream phase.
	\end{itemize}

	\begin{table*}[!t]
		\centering
		\caption{Evaluation Criteria for Manuscript-Level Research Validity.}
		\label{tab:evaluation_rubric}
		\setlength{\tabcolsep}{4pt}
		\renewcommand{\arraystretch}{1.12}
		\scriptsize
		\begin{tabular}{p{0.3\textwidth} p{0.1\textwidth} p{0.42\textwidth}}
			\hline
			\textbf{Dimension} & \textbf{Score} & \textbf{Evaluation focus} \\
			\hline
			Problem definition and scope & 10 & Clarity, boundary, and appropriateness of the research problem. \\
			Novelty and positioning & 10 & Claimed contribution and relation to prior work. \\
			System model and technical correctness & 15 & Correctness of the model, assumptions, notation, and wireless reasoning. \\
			Method validity and formulation alignment & 15 & Consistency between formulation, method, and claimed solution route. \\
			Evidence validity and experiment design & 20 & Execution provenance, baselines, scenarios, sweeps, and empirical support. \\
			Result interpretation and claim support & 15 & Whether numerical trends support the stated claims. \\
			Scientific writing and presentation & 10 & Organization, readability, figure/table quality, and technical clarity. \\
			Reference grounding & 5 & Citation relevance and grounding of prior-work claims. \\
			\hline
			\textbf{Total} & \textbf{100} & -- \\
			\hline
		\end{tabular}
	\end{table*}
	\section{Comparative Evaluation Results}
	\label{sec:results}
	
	In this section, we evaluate the research quality of WARA-generated manuscripts using a structured LLM-based \emph{scoring agent} as the evaluator. The evaluation measures the overall validity of the final manuscript, including problem definition, novelty, technical correctness, method--formulation alignment, evidence support, claim consistency, writing quality, and reference grounding, and we report a manuscript-level research-validity comparison across the considered benchmark sets.
	
	\subsection{ScoringAgent and Evaluation Criteria}
	\label{subsec:scoring_agent}
	
	We design a structured LLM-based {ScoringAgent} to evaluate the research quality of manuscripts. For each evaluated manuscript, ScoringAgent receives a manuscript PDF and calls an evaluator LLM with a fixed reviewer-style prompt. The prompt instructs the LLM to act as a strict wireless-communications reviewer, apply predefined scoring criteria, justify each score using manuscript-visible evidence, and return a machine-readable JSON report. The report contains dimension-level scores, brief justifications, an overall score, identified strengths and weaknesses, and a confidence value. The same scoring prompts, criteria, output schema, parsing rules, and score-range checks are used for all evaluated manuscripts.
	
	The evaluation criteria measure manuscript-level research validity, as summarized in Table~\ref{tab:evaluation_rubric}. They evaluate whether the manuscript defines a clear research problem, positions its contribution, presents a technically correct system model, aligns the method with the formulation, provides valid evidence, supports its claims, communicates clearly, and grounds its statements in references. Together, these criteria define an additive 100-point aggregate quality score over eight dimensions.
	
	For the $i$-th manuscript $\mathcal P_i$, ScoringAgent assigns the dimension scores
	$
	\mathbf s_i^{\mathrm R}
	=
	[s_{i,1}^{\mathrm R},s_{i,2}^{\mathrm R},\ldots,s_{i,8}^{\mathrm R}],
	$
	according to the eight research-validity dimensions in Table~\ref{tab:evaluation_rubric}, and the manuscript-level research-validity score is computed as
	$
	S_i^{\mathrm R}
	=
	\sum_{d=1}^{8}s_{i,d}^{\mathrm R}.
	$
	For a benchmark set $\mathcal B$ with $N$ manuscripts, we report the average score
	$
	\bar S^{\mathrm R}(\mathcal B)
	=
	\frac{1}{N}\sum_{\mathcal P_i\in\mathcal B}S_i^{\mathrm R}.
	$
	\subsection{Benchmark Sets and Comparative Results}
	\label{subsec:benchmark_results}
	
	We compare three benchmark sets. The WARA and one-shot LLM sets are topic-paired, while the accepted IEEE Wireless Communications Letters (WCL) set is used as a peer-reviewed reference profile. Specifically, the benchmark sets are:
	
	\begin{itemize}
		\item \textbf{WARA-generated manuscripts:} ten manuscripts produced by the proposed closed-loop WARA workflow from ten wireless optimization topics, using OpenAI GPT-5.5 as the backbone LLM.
		\item \textbf{One-shot LLM manuscripts:} ten manuscripts generated from the same ten topics using a single prompt and the same OpenAI GPT-5.5 backbone, without phase-structured artifact control, executable validation, or repair.
		\item \textbf{Accepted WCL papers:} ten randomly selected, recently accepted optimization-related IEEE WCL papers used as a peer-reviewed reference group.
	\end{itemize}
	
	To reduce evaluator--generator coupling, the scoring agent is implemented using Kimi K2.6 rather than the GPT-5.5 backbone used for the generated manuscripts.
	
	\begin{table*}[!t]
		\centering
		\caption{Manuscript-Level Validity Scores.}
		\label{tab:research_validity_scores}
		\setlength{\tabcolsep}{4pt}
		\renewcommand{\arraystretch}{1.12}
		\scriptsize
		\begin{tabular}{l c c c c c c c c c c}
			\hline
			\textbf{Benchmark set} & \textbf{$N$} & \textbf{Overall} & \textbf{Scope} & \textbf{Novelty} & \textbf{Model} & \textbf{Method} & \textbf{Evidence} & \textbf{Claims} & \textbf{Writing} & \textbf{Refs} \\
			& & \textbf{/100} & \textbf{/10} & \textbf{/10} & \textbf{/15} & \textbf{/15} & \textbf{/20} & \textbf{/15} & \textbf{/10} & \textbf{/5} \\
			\hline
			One-shot LLM & 10 & 37.4$\pm$2.7 & 6.0 & 5.7 & 7.9 & 6.7 & 0.0 & 3.0 & 5.0 & 3.3 \\
			WARA & 10 & 68.5$\pm$5.7 & 8.0 & 6.8 & 10.4 & 9.8 & 13.0 & 9.3 & 6.9 & 3.6 \\
			Accepted WCL & 10 & 81.4$\pm$4.1 & 8.4 & 7.6 & 12.2 & 11.7 & 17.1 & 12.3 & 8.0 & 4.2 \\
			\hline
		\end{tabular}
	\end{table*}
	
	\begin{figure}[!t]
		\centering
		\includegraphics[width=1\linewidth]{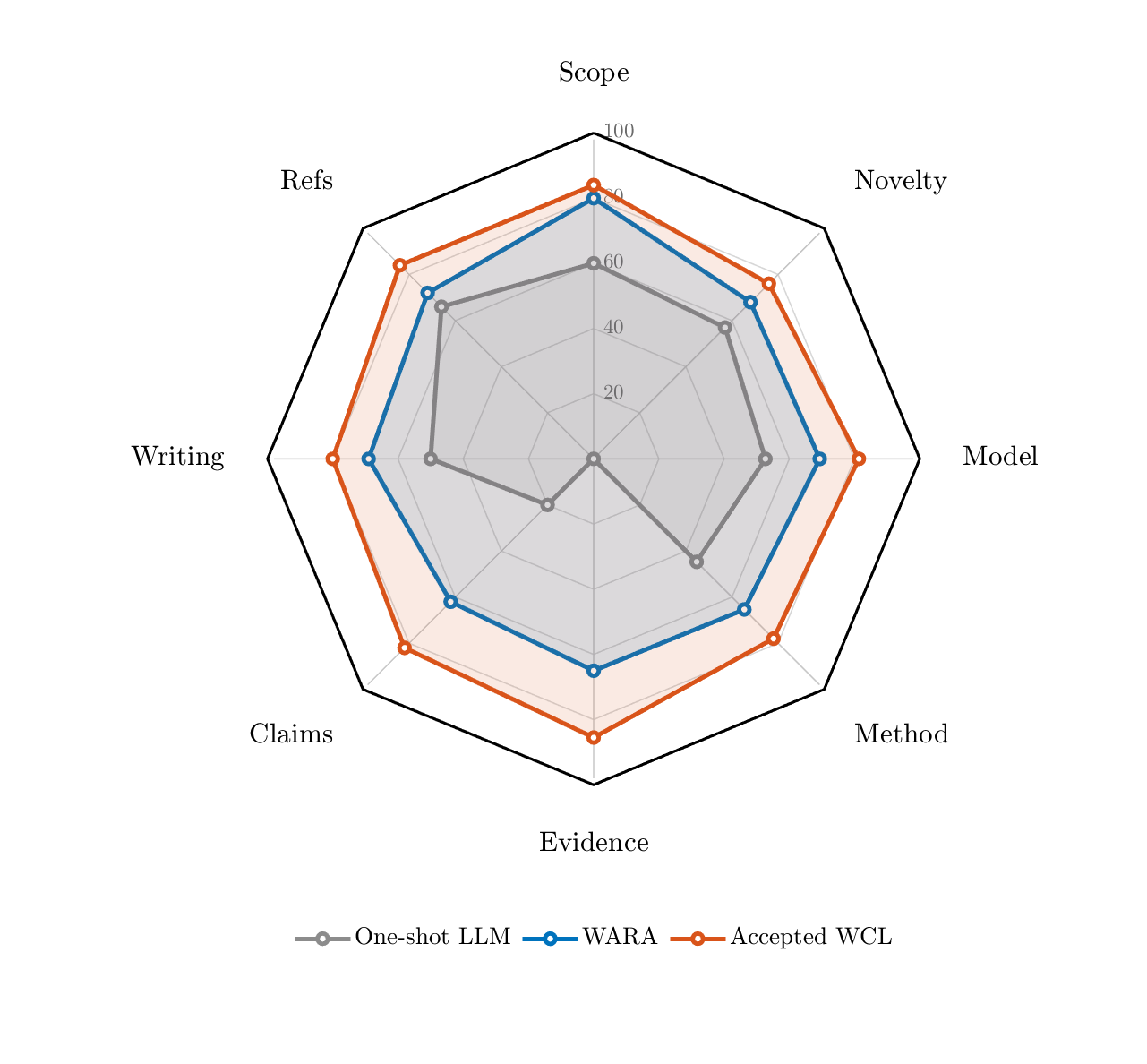}
		\caption{Manuscript-level validity profiles.}
		\label{fig:dimension_profile}
	\end{figure}
	\begin{figure}[!t]
		\centering
		\includegraphics[width=1\linewidth]{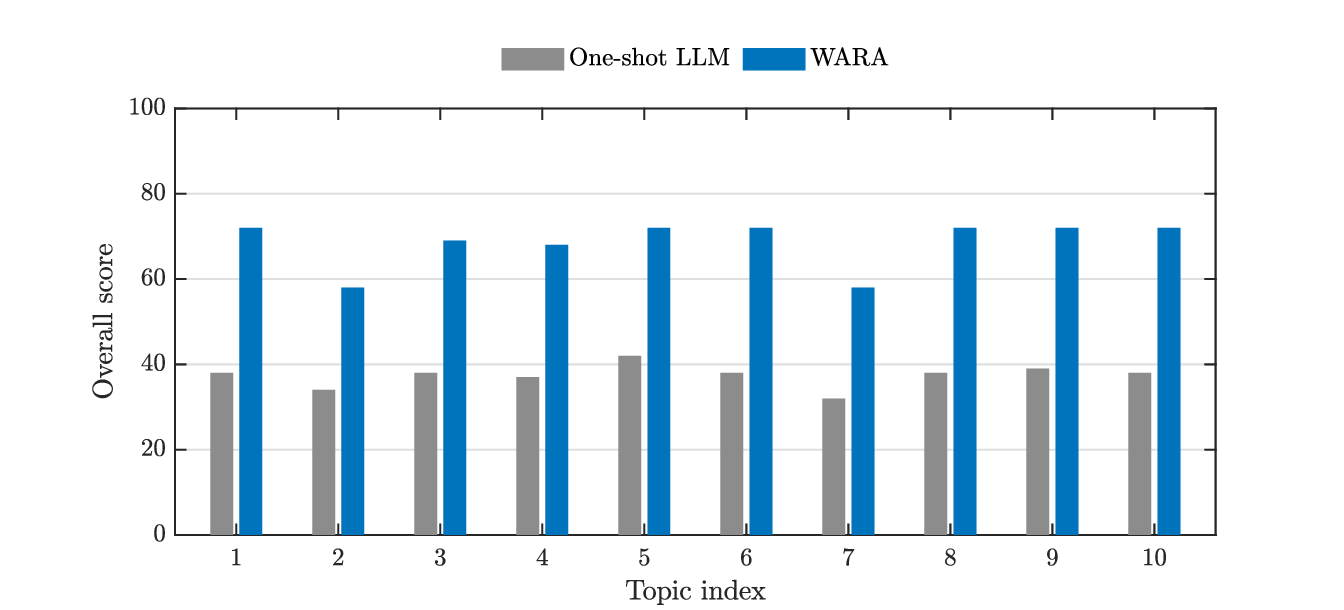}
		\caption{Topic-level paired manuscript scores for WARA and the one-shot LLM baseline.}
		\label{fig:manuscript_scores}
	\end{figure}
	
	Table~\ref{tab:research_validity_scores} and Fig.~\ref{fig:dimension_profile} demonstrate the manuscript-level research-validity comparison. WARA obtains an average score of 68.5, compared with 37.4 for the topic-paired one-shot LLM baseline, yielding a 31.1-point improvement under the same topic set and backbone model. The largest gap appears in evidence validity, where the one-shot baseline receives 0.0 because its numerical results are generated as text without recorded experiment execution, while WARA reaches 13.0 through executable validation and evidence packaging. WARA also improves claim support from 3.0 to 9.3, showing that its performance claims are more directly tied to validated figures and recorded experiment outputs. These two dimensions benefit most from closed-loop control because WARA explicitly separates experiment generation, execution, evidence verification, and claim writing into different gated stages. As a result, unsupported numerical descriptions are blocked before synthesis, and downstream claims must remain within the scope of the verified evidence. The gains in system-model correctness and method validity further show that staged formulation and algorithm construction help reduce mismatch between the stated optimization problem and the proposed solution. Compared with accepted WCL papers, WARA is closer in problem definition and novelty, but still weaker in evidence validity, claim support, and method--formulation alignment. This suggests that WARA improves the reliability of generated manuscripts mainly by making the research process executable, while the remaining gap comes from the depth and refinement of human-designed experiments and technical arguments.
	
	Fig.~\ref{fig:manuscript_scores} shows the topic-level paired comparison between WARA and the one-shot LLM baseline. WARA outperforms the corresponding one-shot manuscript on all ten topics, showing that the gain is not caused by a few favorable examples. The reason is that WARA does not rely on a single forward generation from topic to manuscript. Instead, each topic is repeatedly interpreted, checked, and revised through multiple agent stages: the research direction is reviewed before formulation, the formulation is frozen before algorithm design, the algorithm is checked before implementation, and the evidence is validated before writing. When a gate fails, the repair mechanism sends the problem back to the responsible stage instead of letting the final manuscript absorb the inconsistency. This multi-stage reasoning and targeted repair process explains why WARA produces more stable improvements across different wireless optimization topics.

	\section{Conclusion}
	\label{sec:conclusion}
	
	This paper presented WARA, a closed-loop multi-agent framework for automated wireless optimization research that represents a research study as a chain of verifiable artifacts rather than a single generated manuscript. Starting from an initial topic, WARA identifies a research direction, formulates and solves the corresponding optimization problem, generates executable evidence, and synthesizes the final deliverables, while controller-managed gates, frozen contracts, and targeted repair maintain consistency throughout the workflow. The comparative evaluation shows that WARA substantially outperforms one-shot LLM generation, particularly in formulation--method alignment, evidence validity, and claim support. The remaining gap to accepted peer-reviewed papers mainly lies in experimental depth and evidence maturity. Future work will integrate human expertise directly into the artifact chain through revised contracts, gate criteria, benchmark requirements, simulator settings, and evidence-to-claim rules, enabling expert decisions to be reused across research runs while reserving human intervention for cases that cannot be resolved through automatic validation.

	\bibliographystyle{IEEEtran}
	\bibliography{ref}
	
\end{document}